\documentclass[aps,prl,twocolumn,raggedbottom,showpacs,floatfix,superscriptaddress]{revtex4}

\usepackage{amsmath,amsfonts,amssymb,amsthm,graphics}
\usepackage[next]{inputenc}
\usepackage[dvips]{epsfig}
\usepackage[colorlinks=true,citecolor=blue,linkcolor=blue]{hyperref}
\usepackage{bbm}
\usepackage{bbold}
\usepackage{booktabs}
\usepackage{multirow}
\usepackage{hhline}
\usepackage{amssymb}
\usepackage{comment}

\usepackage{bm}
\usepackage{color}
\usepackage{graphicx,latexsym}
\usepackage{epstopdf}

\usepackage{esvect}
\usepackage{fixmath}

\def\be{\begin{equation}}
\def\ee{\end{equation}}

\def\bi{\begin{itemize}}
\def\ei{\end{itemize}}
\def\bn{\begin{enumerate}}
\def\en{\end{enumerate}}
\def\bea{\begin{eqnarray}}
\def\eea{\end{eqnarray}}
\newcommand{\bpm}{\begin{pmatrix}}
\newcommand{\epm}{\end{pmatrix}}

\def\ba{\begin{array}}
\def\ea{\end{array}}
\def\bd{\begin{displaymath}}
\def\ed{\end{displaymath}}

\renewcommand{\imath}{\hspace{1pt}\mathrm{i}\hspace{1pt}}

\renewcommand{\vec}{\mathbf}

\begin{document}
\title{Amperean pairing at the surface of topological insulators}
\author{Mehdi Kargarian}
\affiliation{Joint Quantum Institute and Condensed Matter Theory Center, Department of Physics, University of Maryland, College Park, MD 20742-4111, USA}
\author{Dmitry K. Efimkin}
\affiliation{Joint Quantum Institute and Condensed Matter Theory Center, Department of Physics, University of Maryland, College Park, MD 20742-4111, USA}
\author{Victor Galitski}
\affiliation{Joint Quantum Institute and Condensed Matter Theory Center, Department of Physics, University of Maryland, College Park, MD 20742-4111, USA}
\affiliation{School of Physics, Monash University, Melbourne, Victoria 3800, Australia}

\begin{abstract}
The surface of a 3D topological insulator is described by a helical electron state with the electron's spin and momentum locked together. We show that in the presence of ferromagnetic fluctuations the surface of a topological insulator is unstable towards a superconducting state with unusual pairing, dubbed Amperean pairing. The key idea is that the dynamical fluctuations of a ferromagnetic layer deposited on the surface of a topological insulator couple to the electrons as gauge fields. The transverse components of the magnetic gauge fields are unscreened and can mediate an effective interaction between electrons. There is an attractive interaction between electrons with momenta in the same direction which makes the pairing to be of Amperean type. We show that this attractive interaction leads to a $p$-wave pairing instability of the Fermi surface in the Cooper channel.   
\end{abstract}
\date{\today}

\maketitle

{\it Introduction.---}
It is known that the Coulomb interaction in normal metals is usually screened by electrons leading to short-ranged and momentum independent interactions, which gives rise to conventional Fermi liquid theory~\cite{landau1,landau2}. The current-current magnetic interaction between electrons, where the interaction is mediated by exchange of transverse photons, however, remains unscreened. Due to the interaction of the gapless bosonic modes with fermions, nonanalytic corrections arise in various physical quantities which
clearly point to non-Fermi liquid behavior~\cite{reizer:prb89, reizer:prb44}. The effect is purely relativistic and is proportional to $(v_{F}/c)^2$, where $v_{F}$ is the Fermi velocity of electrons and $c$ is speed of light. Hence, the corresponding bare interaction is comparatively smaller than coupling constants. 

While the relevance of transverse photons in normal metals and their physical signatures are parametrically small, the search for nonphoton mediated current-current interactions has been extended to other systems such as U(1) gauge fields in the spin liquid description of the Mott phase of organic compounds~\cite{shimizu:prl03,kurosaki:prl05}, a doped Mott insulator~\cite{lee:rmp06} and the Halperin-Lee-Read state~\cite{HLR_state}. The normal state resistivity of a doped Mott insulator exhibits a $T^{4/3}$ temperature dependence~\cite{Plee:prl89} and a $T^{2/3}$ contribution to the specific heat~\cite{Motrunich:prb04,SSlee:prl05} in the presence of the U(1) gauge fluctuations, which manifestly deviates from the Fermi liquid. It was also shown that the U(1) gauge fluctuations can induce a new mechanism for pairing of the spinons in a gapless spin liquid, so-called Amperean pairing~\cite{lee:prl07} with a possible application to the pseudogap phase of cuprate superconductors~\cite{Plee:prx14}.     
 
Given the maturity of heterostructure materials synthesis and recent progress in topological insulators, in this letter we propose a realistic system where the interaction between the fermions and gapless bosons can be engineered to realize an effective fermion-gauge theory and an
Amperean superconductor. The system is made of a ferromagnetic (FM) layer deposited on the surface of a 3D topological insulator (TI) such as Bi$_{2}$Te$_{3}$ and Bi$_{2}$Se$_{3}$. The surface state consists of a single helical Dirac cone~\cite{Xia:np09,Chen:science09}, where the electron spin and momentum are locked together, affecting the transport phenomena and collective excitations~\cite{Raghu:prl10,efimkin:nrl2012}. 

Some previous works have focused on the effect of static~\cite{nomura:prb82, Garate:prl10,Tserkovnyak:prl108, Ferreiros:prb89,hurst:prb91,Hurst:prb16} and dynamical~\cite{Yokoyama:prb10, cenke:prb10, nogueira:prb88,potter:prb15} magnetic fluctuations. We consider the effect of transverse dynamical magnetic fluctuations on a doped Dirac cone. We will show that the latter have profound effects on helical states. Here is a summary of our results: (i) transverse magnetic fluctuations are unscreened and mediate an effective interaction between electrons; (ii) the effective interaction is singular at small frequency and momentum transfer and leads to non-Fermi liquid behavior at very low energies;  (iii) the effective interaction has an Amperean form: it is attractive between electrons near the Fermi surface moving in the same direction; (iv) the attractive interaction leads to a pairing instability of the Fermi surface, dubbed Amperean pairing.  


\begin{figure*}[!htb]
\includegraphics[width=15cm]{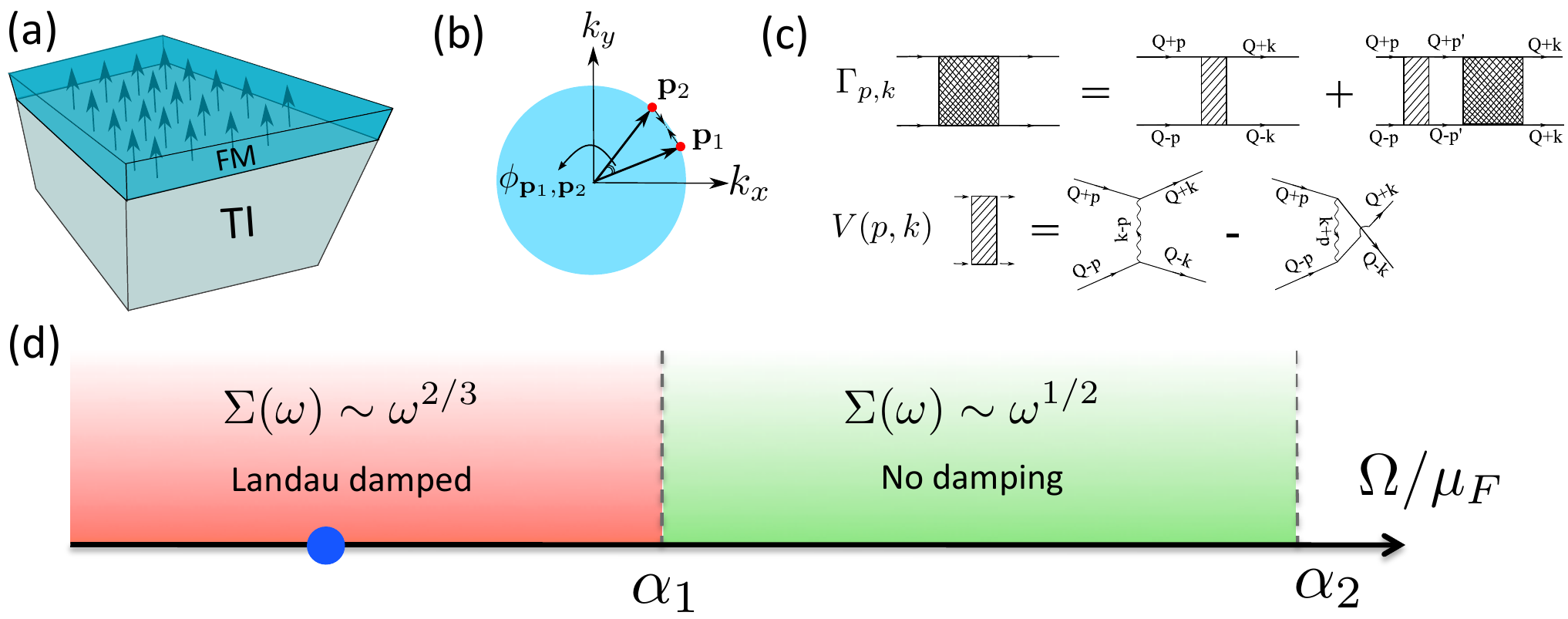}
\caption{(color online) (a) A ferromagnetic (FM) layer deposited on the surface of a topological insulator (TI), (b) the blue (light) disk indicates the region in momentum space bounded by a Fermi surface with Fermi energy $\mu_{F}$. Small arrows on the Fermi surface indicate an attractive interaction between electrons (red balls) with momenta $\mathbf{p}_{1}$ and $\mathbf{p}_{2}$ and the angle between them is $\phi_{\mathbf{p}_{1},\mathbf{p}_{2}}=\phi_{\mathbf{p}_{2}}-\phi_{\mathbf{p}_{1}}$, (c) Feynman diagrammatic representation of Bethe-Salpeter equation and the interaction in direct and exchange channels, and in (d) we show the regions with the non-Fermi liquid behavior and pairing instability at zero temperature. The horizontal axis stands for the frequency of transverse magnon excitations with propagator $D_{T}(q)$. In the low frequency regime, the excitations are highly damped and $\Sigma(\omega)\sim\omega^{2/3}$, while in the opposite regime up to an energy scale given by $\alpha_{2}$ the magnons are described by a bare propagator $D^{0}_{T}(q)$ and $\Sigma(\omega)\sim\omega^{1/2}$ dominates over the linear term. In both regimes the system is unstable to Amperean pairing. For some experimentally relevant parameters the propagator of magnons falls into the damped regime shown by a blue solid disk with strong instability to Amperean pairing.}\label{fig_prl}
\end{figure*}        

{\it Model.---}
Consider a hybrid system of a ferromagnet with spin density $\rho_{s}$ and the surface of TI as shown in Fig. \ref{fig_prl}(a). The continuum action of a quantum ferromagnet with local moments $\mathbf{S}=S\mathbf{n}$ is given by~\cite{Fradkin} 
\bea \mathcal{S}_{m}=\frac{\rho_{s}}{2}\int d\tau d^{d}x\left[-2i\mathbf{A}(\mathbf{n})\cdot\partial_{\tau}\mathbf{n}+\kappa\left(\nabla\mathbf{n}\right)^2\right].\eea 
Here, $\kappa=JSa_{0}^2$, where $J$ is the ferromagnetic exchange coupling and $a_{0}$ is the lattice spacing. We use $\kappa$ to define magnon mass $m_\mathrm{s}$ via $\kappa=1/2m_{s}$ ($\hbar=1$) below. The first term in the action is the Wess-Zumino term where the vector potential $\mathbf{A}$ creates a local magnetic field as $\nabla_{\mathbf{n}}\times\mathbf{A}(\mathbf{n})=\mathbf{n}$. We assume that the magnetic order in the ferromagnet is along the z-direction $\mathbf{z}$. In order to describe the ferromagnetic magnons $\vec{l}(\tau,\vec{r})$, we write $\vec{n}=\vec{z}\sqrt{1-|\mathbf{l}|^2}+\vec{l}$, where $|\mathbf{l}|\ll 1$ and $\vec{z}\cdot\vec{l}=0$, leading to $\mathbf{A}=1/2(\vec{z}\times\mathbf{l})$. Neglecting orders higher than quadratic terms, the action of magnons is written as~\cite{Nayak}      
\bea \label{Smagnon} \mathcal{S}_{m}=\frac{\rho_{s}}{2}\int d\tau dx^d \left[i(\vec{l}\times \partial_{\tau}\vec{l})_z+\kappa(\nabla  \vec{l})^2\right].\eea

The following action describes the Dirac electrons and their coupling to the ferromagnet moments: 
\bea \mathcal{S}_{D}=\int {d\tau d\vec{r}} \bar{\psi}\left[\partial_{\tau}+v_{F} (\vec{p}\times \mathbold{\sigma})_z-\mu_{F}-g\vec{n} \cdot \mathbold{\sigma}\right]\psi. 
\eea The spinor fields $\psi(\tau,\vec{r})=(\psi_{\uparrow},\psi_{\downarrow})^T$ describe the electrons and $\mathbold{\sigma}$ is a vector of the Pauli matrices, representing spin. Other parameters are the Fermi energy $\mu_{F}$ and the coupling $g$ between electrons and magnetic moments. As we concentrate on the doped regime which is relevant to the experiments~\cite{Xia:np09,Chen:science09}, we will ignore a uniform coupling $g \sigma_z$ to the electrons that opens up a gap of magnitude $2|g|$ at the Dirac node.

The helical nature of surface states allows us to present the magnetic fluctuations as dynamical gauge fields $\vec{a} = g v_{F}^{-1}~\vec{l}\times\mathbf{z}$  minimally coupled to electrons. It can be split into longitudinal and transverse components as $\vec{a} = \vec{a}_L + \vec{a}_T$ so that $\nabla\cdot\mathbf{a}_{T}=0$ and $\nabla\times\mathbf{a}_{L}=0$. The transverse part is responsible for the emergent magnetic field $B_z=(\nabla \times \vec{a}_T)_z=-g v_{F}^{-1} \nabla\cdot\vec{l}_L$ perpendicular to the surface, and the longitudinal part generates an emergent electric field $\vec{E}=-\partial_t  \vec{a}_L=-g v_{F}^{-1}\partial_t \vec{l}_T$. Here, $\vec{l}_L$ and $\vec{l}_T$ are, respectively, the longitudinal and transverse components. A local gauge transformation can eliminate the longitudinal part of the gauge field $a_L$ and transforms it to the scalar potential $\varphi$, i.e. the temporal component. The resulting action, describing electrons coupled to the transverse
gauge field and the scalar potential, can be written as follows.
\begin{equation}
\mathcal{S}_{D}=\int d\tau d\textbf{r}\bar{\psi}\left[\partial_{\tau}+v_{F}\left(\mathbold{\Pi}\times \mathbold{\sigma}\right)_z-\mu_{F}+ i \varphi \right]\psi, 
\end{equation} where $\mathbold{\Pi}=\vec{p}-\vec{a}_T$.

{\it Bosonic propagators.---} We express the bare magnon action in Eq.~(\ref{Smagnon}) in terms of transverse and temporal components. We obtain the temporal $D^{0}_{\varphi}(q)=\langle\varphi(-q),\varphi(q)\rangle$ and transverse $D^{0}_{T}(q)=\langle a_{T}(-q),a_{T}(q)\rangle$ propagators by, respectively, integrating out the transverse and temporal fields as follows.  
\bea D^{0}_{\varphi}(q)=\frac{1}{2\rho_{s}}\frac{2\kappa q_{n}^2}{q_{n}^2+\Omega_{\mathbf{q}}^2},~~D^{0}_{T}(q)=\frac{1}{2\rho_{s}}\frac{2\Omega_{\mathbf{q}}}{q_{n}^2+\Omega_{\mathbf{q}}^2}. \eea
Here, $q=(q_{n},\mathbf{q})$, where $q_{n}=2n\pi/\beta$ with $\beta=1/T$, and $\mathbf{q}$ is 2d momentum. We use real frequencies $\Omega$ and $\omega$ for analytically continued boson and fermion propagators, respectively, throughout. The magnon dispersion is $\Omega_{\mathbf{q}}=\kappa|\mathbf{q}|^2$. As we are interested in the low frequency regime, where the transverse field propagator is singular and the Amperean pairing sets in, the propagator for temporal fields is not singular enough and can be ignored. Therefore, the pairing instability sets in even in the presence of a repulsive interaction due to the temporal components~\cite{Plee:prl89}. We are left with a theory of helical electrons coupled to the massless and unscreened transverse magnons. We write the latter in terms of bare and dressed magnon propagators characterized by a dimensionless coupling constant $\alpha_0=g^2\rho_F/\rho_s \Omega_s=u\eta^{-1} $, where $\rho_{F}$ is the density of states at the Fermi surface and $\Omega_s=\eta  \mu_F$ is to be understood as a maximum energy transferred by magnons. Here, we introduced two dimensionless controlling parameters $u=g^2\rho_F/\rho_s \mu_F$ and $\eta=m_D/m_s$, where $m_D=\mu_{F}/v_{F}^2$ is to be interpreted as the mass of Dirac electrons. For a typical set of parameters, which we discuss at the end of the paper, we have $\alpha_0\sim 1$ and $\eta\ll 1$. Hence, not only does the strength of the Amperean interaction becomes singular especially near the bosonic poles, but it also acquires moderate strength even away from the bosonic poles. The presence of a small parameter $\eta\ll 1$, which implies that magnons are much slower than Dirac fermions, allows us to neglect vertex corrections in the spirit of the Migdal theorem~\cite{migdal,roy:prb89}. As a result, the dressed propagator of transverse fields in the one-loop approximation is given by
\bea \label{DT} D_{T}(q)=\frac{1}{2\rho_{s}} \frac{2 \Omega_{\mathbf{q}} }{\Omega_{\mathbf{q}}^2+q_{n}^2+\gamma_{0} \Omega_{\mathbf{q}} \frac{|q_{n}|}{v|\mathbf{q}|}},\eea
where $\gamma_{0}=g^2\rho_{F}/\rho_{s}$ is the Landau damping. In the regime $\Omega\gg \alpha_{1} \mu_{F}$, where $\alpha_{1}\simeq\alpha_{0}\eta^2$, the Landau damping term in the denominator of Eq.~(\ref{DT}) can be ignored reducing the dressed propagator $D_{T}(q)$ to the bare propagator $D_{T}^{0}$. On the other hand for $\Omega\ll \alpha_{1} \mu_{F}$ the Landau damping term dominates and the propagator becomes $D_{T}(q)=1/\rho_{s}(\kappa|\mathbf{q}|^2+\gamma_{0}|q_{n}|/v|\mathbf{q}|)$ resembling the critical bosonic soft modes $\Omega\sim q^z$ with dynamical exponent $z=3$ appearing in different contexts such as gauge theory \cite{Ioffe:prb89} of high-$T_{c}$ superconductors \cite{Lee:prl89_HTc} and critical ferromagnetic systems \cite{chubukov:prb06}. There are also other energy scales $\omega\sim \alpha_{2,3}\mu_{F}$, where $\alpha_{2}\simeq\alpha_{0}^2\eta$ and $\alpha_{3}\simeq\alpha_{0}^2$, which determine the non-Fermi liquid behavior of electrons. We will discuss this below.  

{\it Transverse magnon-mediated interaction.---}
Upon integrating out the transverse magnons and projecting into the conduction band, we obtain the effective interaction between electrons as
\bea\label{Sint} \mathcal{S}_{int}=-\frac{1}{2\mathcal{V}\beta} \sum_{p_{1},p_{2},q} V_{\mathbf{p}_{1},\mathbf{p}_{2}}(q)\bar{\psi}_{p_{1}+q}\bar{\psi}_{p_{2}-q}\psi_{p_{2}} \psi_{p_{1}},\eea 
where $\mathcal{V}$ is the volume of the system. Note that the fermionic fields $\psi$'s have only a conduction band index, so the interaction is between the effective spinless fermions. The key observation in Eq.~(\ref{Sint}) is the form of the interaction: that is $V_{\mathbf{p}_{1},\mathbf{p}_{2}}(q)=g^2D_{T}(q)\Lambda_{\mathbf{p}_{1},\mathbf{p}_{2}}(\mathbf{q})$ with 
\bea \label{Lambda} \Lambda_{\mathbf{p}_{1},\mathbf{p}_{2}}(\mathbf{q})=\frac{1}{2}\left[\cos(\phi_{\mathbf{p}_{1}}-\phi_{\mathbf{p}_{2}})-\cos(\phi_{\mathbf{p}_{1}}+\phi_{\mathbf{p}_{2}}-2\phi_{\mathbf{q}})\right],\eea 
where $\phi$'s are the angles characterizing the corresponding momentum. For small momentum transfers $|\mathbf{q}|\ll |\mathbf{Q}|$, where $\mathbf{Q}$ is the Fermi momentum, the angle $\phi_{\mathbf{q}}$ varies between zero and $2\pi$ for fixed incoming momenta, making the second term vanish in angular integration on $\phi_{\mathbf{q}}$. Thus, $\Lambda$ depends upon the angle between two incoming momenta: $\Lambda\simeq \hat{p}_{1}\cdot\hat{p}_{2}/2$, where the hat stands for unit vector. Importantly, $\Lambda$ is positive for electrons moving in almost the same direction, making the interaction attractive. This resembles attraction between two wires carrying co-moving currents and is the basis of the Amperean pairing theory as
originally proposed by Lee {\it et al}~\cite{lee:prl07}. Hence, the source of attractive interaction in our case is markedly different from other spin fluctuation mediated pairings in, e.g., pnictides~ \cite{FaWang:science11,zhuang:prb11} or UCoGe~\cite{Hattori:prl12}.

The attractive interaction between electrons can lead to a Cooper-pairing instability that we explore below. Unlike the conventional superconductors with singlet pairing between electrons residing on opposite sides of the Fermi surface, Amperean pairing occurs between electrons residing on the same side of the Fermi surface. This means that in Eq.~(\ref{Sint}) the incoming momenta are to be taken as $\mathbf{p}_{1}=\mathbf{Q}+\mathbf{p}$ and $\mathbf{p}_{2}=\mathbf{Q}-\mathbf{p}$ with $|\mathbf{p}|\ll |\mathbf{Q}|$. This leads to $\Lambda\simeq 1/2$. In examining the pairing instability, the interactions should be treated in both direct and exchange channels \cite{coleman:book} due to the spinless interaction. This amounts to rewriting the interaction as 
\bea \label{Sint_p} \mathcal{S}_{int}=-\frac{1}{2\mathcal{V}\beta}\sum_{p,k}V(k,p)\bar{\psi}_{Q+k}\bar{\psi}_{Q-k}\psi_{Q-p} \psi_{Q+p},\eea
where $V(k,p)=1/4\left[D_{T}(k-p)-D_{T}(k+p)\right]$ and $k=q+p$ as shown diagrammatically in Fig.~\ref{fig_prl}(c). Given the $p$-wave character of the pair wavefunction, both channels have the same contributions to the eigenvalue problem below.

{\it Amperean pairing.---}
In order to examine the Amperean pairing instability of the Fermi surface, we examine the Bethe-Salpeter equation for the effective interaction in Eq.~(\ref{Sint_p}) in the Cooper channel as shown diagrammatically in Fig. \ref{fig_prl}(c).  
\bea \label{BS} \Gamma_{p,k}=V(p,k)+\frac{1}{\beta}\sum_{p'}V(p,p')G_{Q}(p')G_{Q}(-p')\Gamma_{p',k},\eea
where $G_{Q}(\pm p)$ is the dressed Green function of electrons with energy dispersion $\varepsilon_{\mathbf{Q}\pm\mathbf{p}}$.   
Since only one of the momentum components of $\Gamma_{p,k}$ is involved in the sum, it amounts to an eigenvalue problem. Lets write $\Gamma_{p,k}=\Phi_{k}(p)$ and re-cast it into
\bea \label{eigenvalue} E\Phi_{k}(p)=\frac{1}{\beta}\sum_{p'}V(p,p')G_{Q}(p')G_{Q}(-p')\Phi_{k}(p'). \eea 
This is the Bardeen-Cooper-Schrieffer (BCS) self-consistency equation and $\Phi$ measures the pairing amplitude \cite{Galitski:prl07}. The instability towards Amperean pairing is signaled as $E\geq1$. We shall argue that the strength of singularity in the Cooper channel depends on the renormalization of electrons by transverse magnons via the electron self-energy $\Sigma(p)$.  

We now elaborate on the fermionic self-energies in the regimes where $\Omega\ll \alpha_{1}\mu_{F}$ and $\Omega\gg \alpha_{1}\mu_{F}$; see also Fig.~\ref{fig_prl}(d). In the former, the electron propagator is given by $G(p)=1/[i(p_{n}+\epsilon_{L}|p_{n}/\mu_{F}|^{2/3}\mathrm{sgn}(p_{n}))-\varepsilon(\mathbf{p})]$, where $\epsilon_{L}=\alpha_{0}^{2/3}\mu_{F}/2\sqrt{3}$ and $\varepsilon(\mathbf{p})$ is the electron dispersion on the surface of the TI. Note that we considered the frequency dependent self-energy correction near the Fermi surface. The momentum part only renormalizes the mass of the electrons. It turns out that in this regime the non-analytic term in the self-energy is much larger than $\omega$ in the denominator of the electron propagator. Hence, the system becomes a non-Fermi liquid at low energies $\omega\ll \alpha_{3}\mu_{F}$ where $\alpha_{3}\simeq\alpha_{0}^2$. Note that because $\alpha_{3}/\alpha_{1}\simeq \alpha_{0}\eta^{-1}\gg 1$, the non-Fermi liquid behavior persists even up to high frequency transfer of magnons.   

On the other hand, for the latter regime where $\Omega\gg\alpha_{1}\mu_{F}$, the transverse magnons are not substantially Landau damped. In this case the electron propagator is dressed differently and is given as $G(p)=1/[i(p_{n}+\epsilon_{N}|p_{n}/\mu_{F}|^{1/2}\mathrm{sgn}(p_{n}))-\varepsilon(\mathbf{p})]$, where $\epsilon_{N}=\alpha_{0}\eta^{1/2} \mu_{F}/8\pi$. The self-energy could be parametrically larger than $\omega$ for $\omega\ll\alpha_{2}\mu_{F}$ where $\alpha_{2}\simeq\alpha_{0}^2\eta$. Note that $\alpha_{1}\ll\alpha_{2}$. Hence, there exists a set of energy scales $\alpha_{1}\ll\alpha_{2}\ll\alpha_{3}$ determining the behavior of boson and electron propagators. In treating the eigenvalue equation in Eq.~(\ref{eigenvalue}), we focus on low energy limits of non-Fermi liquids set by $\alpha_{1,2}$ as shown in Fig.~\ref{fig_prl}(d): (i) Landau damped and (ii) non-Landau damped regimes. In both regimes the system is unstable to Amperean pairing. We discuss each regime separately.

(i) In the Landau damped regime the dressed propagators are analogous to spinons coupled to singular gauge fields \cite{lee:prl07}. At zero temperature we replace the Matsubara sum with an integral over $p_{0}$ and the momentum integral over $\mathbf{p}$ is written as two one-dimensional integrals over $p_{\perp}$ and $p_{\parallel}$ where $p_{\perp}$ ($p_{\parallel}$) is the momentum component perpendicular (parallel) to the Fermi momentum. Following the ansatz presented in Ref. [\onlinecite{lee:prl07}] for the wavefunction $\Phi(\mathbf{p})=\tilde{\Phi}(p_{\perp})\Theta(p^2_{\perp}/Q-|p_{\parallel}|)$, where $\Theta$ is the Heaviside function, the eigenvalue equation becomes   
\bea \label{Ephi} E\tilde{\Phi}(p_{\perp})=\int dt K(p_{\perp},t)\tilde{\Phi}(t+p_\perp),\eea 
where the kernel $K(p_{\perp},t)$ is given by
\bea \label{kernel_dam} K(p_{\perp},t)= \frac{|t|}{\sqrt{3}\pi(t+p_{\perp})^2} \ln\left[\frac{t^4/3+9(t+p_{\perp})^4}{t^4/3+(t+p_{\perp})^4} \right]. \eea 

To get an insight into $E$, the first approximation would be to consider a momentum independent wavefunction. The rest of the integral is logarithmically diverging, which signals the possibility of pairing. The existence of a realistic and nontrivial solution, however, requires that the momentum dependent wavefunction is taken into account. Our numerical calculations and the results presented in Ref. \cite{lee:prl07} show that there exists such a solution with odd pairing wavefunction, and the corresponding eigenvalue becomes larger than unity for a large enough system. Therefore, the system is unstable to Amperean pairing. Interestingly enough, the magnetic coupling $g$ doesn't appear explicitly in Eq.~(\ref{kernel_dam}). Thus, the Amperean pairing sets in even at small couplings, so long
as the bosonic propagator is in the highly damped regime as shown in
Fig.~\ref{fig_prl}(d). Indeed, the instability at zero temperature here is due to the
fact that the kernel is highly singular. This result indicates that
Amperean pairing occurs even at finite temperatures, which is the main
experimentally-relevant conclusion of this work.   

(ii) In the non-Landau damped regime we found the eigenvalue equation to be less singular. We obtained the following expression for the kernel in Eq.~(\ref{Ephi}). 
\bea K(p_{\perp},t)= \frac{A}{4\sqrt{2}\pi(t+p_{\perp})^2} \ln\left[\frac{A^2t^2+9(t+p_{\perp})^4}{A^2t^2+(t+p_{\perp})^4} \right],\eea
where $A=\mu Q$. The eigenvalue problem contains no dimensional parameters. Upon changing variable $t\rightarrow t-p_{\perp}$ and introducing dimensionless variables $x=t/A$ and $y=p_{\perp}/A$, the corresponding eigenvalue equation can be written as follows.
\bea E\tilde{\Phi}(y)=\frac{1}{4\sqrt{2}\pi} \int\frac{dx}{x^2} \ln\left[\frac{(x-y)^2+9x^4}{(x-y)^2+x^4} \right]\tilde{\Phi}(x). \eea
The natural ansatz to check is a constant wavefunction $\tilde{\Phi}(x)={\rm const}.$ at $y=0$ corresponding to pairing between electrons right at the Fermi surface. This gives rise to $E=1$ in contrast to the logarithmic divergence found in Landau damped case. This result shows that the system is quantum critical on the verge of
the Amperean instability. We need, however, to look for a wavefunction which is odd in momentum, i.e. a $p$-wave one.  Viewing the kernel as a matrix, we observed that there exists such a solution. In fact, the momentum dependent solution shows that the maximum eigenvalue of the kernel becomes even larger than unity signaling the instability of the system towards the pairing formation. Therefore, even in the regime with small energy transfer of magnons, as shown in Fig.~\ref{fig_prl}(d), the non-Fermi liquid behavior and pairing instability take place.     

{\it Concluding remarks.---}
In conclusion, we demonstrated that there exists an Amperean pairing instability at the surface of a 3D topological insulator when the Dirac electrons are coupled to ferromagnetic fluctuations. The key idea is that the ferromagnetic fluctuations are minimally coupled to the electrons as gauge fields. The transverse components of the gauge field remains gapless even in the presence of finite chemical doping; they are not screened and mediate an attractive interaction between electrons moving almost in the same direction. 
We showed that there exists strong pairing instability and the system becomes unstable toward a superconducting state with finite-momentum Cooper pairs. The latter state yields some analogy with Fulde-Ferrell-Larkin-Ovchinnikov state~\cite{Fulde-Ferrel,Larkin-Ovchinnikov}, where a non-uniform ground state appears due to completely different reasons. The non-uniform ground state can then be probed by various methods such as scanning tunneling spectroscopy~\cite{Akbari:NJP16}, the Josephson effect~\cite{Yang:prl00,Efimkin:FFLO} and possibly Andreev reflection~\cite{Beenakker:prl15}. For estimations of energy scales we used the following typical set of experimentally relevant parameters $g=25\; \hbox{meV}$ \cite{Chen:science10,Chang:science13,Mingda:prl15,Mingda:prb15}, $\rho_s=4\times10^{12}\; \hbox{cm}^{-2}$, $\mu_F=0.1\;\hbox{eV}$ and $m_s=3.6\times10^{10}\; \hbox{eV} \hbox{s}^2/\hbox{m}^2 $ leading to values of $\alpha_{0}\sim 1$, $u \approx 1.5\times10^{-3}$ and $\eta \approx 10^{-3}$ for dimensionless parameters. For this set of parameters, we found the characteristic energy transfer of magnon excitations falls into the Landau damped regime, the case (i), as marked by a blue disk in Fig.~\ref{fig_prl}(d) with a transition temperature of about $T_{c}\sim 1\;\hbox{K}$. Nevertheless, we also explored the undamped regime, the case (ii), to emphasize the possible extension of pairing instability by tuning the parameters so that the latter regime can be achieved. Two modifications can be made in our system: instead of heterostructure shown in Fig.~\ref{fig_prl}(a) one could also use TI/FM even with a metallic ferromagnet such as Bi/Ni~\cite{BiNi_pwave15}, where a transition to superconductivity at $4\;\hbox{K}$ was reported. The main prediction of this work is that heterosructures (e.g., Bi$_2$Se$_3$ and Bi$_2$Te$_3$ as TI and EuS and Ni as ferromagnets) straightforwardly achievable
with current experimental capabilities~\cite{Peng:prl13,Mingda:prl15}, host  a non-Fermi liquid state
and an exotic Amperean superconductor. \\      

{\it Acknowledgments.---} The authors are grateful to Boris Altshuler and Hilary M. Hurst for valuable discussions and comments. This research was supported by DOE-BES DESC0001911 and the Simons Foundation.

%

\end{document}